\documentclass[preprint, amsmath,amssymb,aps,pra]{revtex4-2}

\usepackage{graphicx}
\usepackage{dcolumn}
\usepackage{bm}
\usepackage{booktabs}
\usepackage{rotating}

\begin{document}

\title{ Improvements in the orbitalwise scaling down of Perdew-Zunger self-interaction correction in many-electron regions}
\author{Yoh Yamamoto}
\email{yyamamoto@utep.edu}
\author{Selim Romero}
\author{Tunna Baruah}
\author{Rajendra R. Zope}
 \email{rzope@utep.edu}
\affiliation{Department of Physics, University of Texas at El Paso, El Paso, Texas 79968}

\date{April 10, 2020}

\begin{abstract}

The Perdew-Zunger (PZ)
method provides a way to remove the self-interaction (SI) error from density functional approximations on an orbital by orbital basis. 
The PZ method provides significant improvements
for the properties such as barrier heights or dissociation energies but results in
over-correcting the properties well described by SI-uncorrected semi-local functional.
One cure to rectify the over-correcting tendency is to scale down the magnitude of 
SI-correction of each  {\bf{orbital}} in the many electron region. 
We have implemented the orbitalwise scaled down SI-correction (OSIC) scheme of 
Vydrov \textit{et al.}  [J. Chem. Phys. {\bf 124}, 094108 (2006)] 
using the Fermi-L\"owdin SI-correction method. 
After validating the OSIC implementation 
with  previously reported OSIC-LSDA results, we examine its
performance with the most successful non-empirical SCAN meta-GGA functional.
Using different forms of scaling factors to identify one-electron regions,
we assess the performance of OSIC-SCAN for a 
wide range of properties: total energies, ionization potentials and 
electron affinities for atoms, atomization energies, dissociation and 
reaction energies, and reaction barrier heights of molecules. 
Our results  show that  OSIC-SCAN provides superior results than the previously reported
OSIC-LSDA, -PBE,  
and -TPSS results.
Furthermore, we propose selective scaling of OSIC (SOSIC) to remove its major shortcoming that  destroys
the $-1/r$ asymptotic behavior of the potentials.
The SOSIC method gives the highest occupied orbital eigenvalues practically identical to those in PZSIC and 
unlike OSIC provides bound atomic anions even with larger powers of scaling factors. 
SOSIC compared to PZSIC or OSIC provides more balanced description
of total energies and barrier heights.
\end{abstract}

\maketitle

\section{\label{sec:introduction}Introduction}
The Kohn-Sham (KS) formulation\cite{PhysRev.140.A1133,PhysRev.136.B864} of the density functional theory (DFT)
is formally an exact approach to obtain the ground state energy
of many electron system.  It is by far the most widely used 
method for obtaining the electronic and structural properties 
of molecules and solids. Its practical applications require 
approximation to the exact exchange-correlation functional 
that is representative of the non-classical energy contributions.
There is no systematic way to construct the functional, and 
a large number of approximate functionals have been
proposed and widely used.
The exchange-correlation functional 
approximations are classified by Perdew and Schmidt\cite{perdew2001jacob} using an analogy to the 
Jacob's ladder wherein the functional approximation corresponds to the rungs
of a ladder. The earliest functional approximation is the
celebrated local spin density approximation (LSDA)\cite{PhysRevB.45.13244} which  forms the first rung of the ladder.
The functionals with more complex ingredients such as density gradients,
density Laplacians or Kohn-Sham orbitals belong to the higher rungs. 
Thus, the generalized-gradient approximation (GGA)\cite{PhysRevLett.77.3865,PhysRevLett.78.1396}
goes beyond the LSDA by 
capturing non-homogeneity of density using density gradients corresponds to the second 
rung. Likewise, the third rung of the ladder corresponds to the 
meta-GGAs that use kinetic energy densities or density Laplacians while the fourth one corresponds
to the hyper-GGA functionals, examples of which are the hybrid functionals\cite{doi:10.1063/1.464304} 
that include certain percentages of the Hartree-Fock (HF) exchange in the functional
approximation. The functionals from the second to fourth rungs (GGAs, meta-GGAs, and hyper-GGAs)
are widely used today in the molecular physics,
solid state physics, and materials science. These functionals can describe many physical properties with sufficient accuracy. Their efficient numerical
implementations, available in a large number of easy-to-use codes, have led to a
proliferation of density functional based studies.
    One shortcoming of the majority of the density functional approximations (DFAs) mentioned above is that these 
    approximations suffer from the self-interaction error (SIE), which arises 
    due to incomplete cancellation of the classical Coulomb interaction of an electron with itself by
    the approximate exchange-correlation term in the energy functional.
    In general, the modern 
    semi-local functionals are sophisticated enough to provide a fairly accurate description
    of the equilibrium properties such as atomization energies but 
    they fail to describe the properties such as 
    transition states in chemical reactions, charge-transfer excitations, 
    binding of an electron in some anions, dissociation of molecules. The
    SIE in these functionals is considered to be responsible for these failures\cite{doi:10.1063/1.1630017,PhysRevB.23.5048}.
    Indeed, the SIE is recognized to be a major limitation of DFAs
    that limits their universal usage\cite{zhang1998challenge,doi:10.1063/1.2741248,ranasinghe2017does,klupfel2013koopmans,gidopoulos2012constraining}.
    In 1981, Perdew and Zunger\cite{PhysRevB.23.5048} (PZ) proposed a method to eliminate
    SIE on an orbital by orbital basis. They applied this self-interaction 
    correction (SIC) scheme to the LSDA which was the only known approximation
    at that time and found significant improvements in atomic properties.
    Their scheme later became known as PZSIC.
    Subsequent calculations on molecules in mid-eighties by the Wisconsin group\cite{PhysRevB.28.5992,doi:10.1063/1.446959,doi:10.1063/1.448266} 
    used localized orbitals to compute the self-interaction (SI) energies of molecules.
    Since then a number of studies have used SIC implementations\cite{ doi:10.1063/1.481421, doi:10.1063/1.1327269, doi:10.1063/1.1370527, harbola1996theoretical, doi:10.1063/1.1468640, doi:10.1021/jp014184v,PhysRevA.55.1765,doi:10.1080/00268970110111788, Polo2003, doi:10.1063/1.1630017, B311840A,doi:10.1063/1.1794633, doi:10.1063/1.1897378, doi:10.1063/1.2176608, zope1999atomic, doi:10.1063/1.2204599,doi:10.1002/jcc.10279,PhysRevA.45.101, PhysRevA.46.5453,lundin2001novel, PhysRevA.47.165,doi:10.1021/acs.jctc.6b00347,csonka1998inclusion,petit2014phase,kummel2008orbital,schmidt2014one,kao2017role,schwalbe2018fermi,jonsson2007accurate,rieger1995self,temmerman1999implementation,daene2009self,szotek1991self,messud2008time,messud2008improved,doi:10.1063/1.1926277,korzdorfer2008electrical,korzdorfer2008self,ciofini2005self,PhysRevA.50.2191,
    doi:10.1063/1.5125205,C9CP06106A,doi:10.1002/jcc.25767,doi:10.1021/acs.jctc.8b00344,doi:10.1063/1.4947042,schwalbe2019pyflosic,
    doi:10.1063/1.4996498, doi:10.1063/1.5050809, doi:10.1021/acs.jpca.8b09940,Jackson_2019, FLOSIC_WATER_PNAS}
    to study atoms, molecules, and solids. 
    It has been found in a number of studies that the PZSIC 
    when used to compute thermochemical properties such as 
    enthalpies of formation provides improvement over the LSDA functional,
    but the results are still not as accurate as those obtained using
    the GGAs. In particular, PZSIC when used with GGAs and meta-GGAs 
    often worsens the results for thermochemical properties. It, however,
    does provide significantly improved results for properties such 
    as reaction barriers and barrier heights where chemical bonds are stretched. 
    This improvement is observed for all the DFAs (LSDA, GGA, and meta-GGAs).
    This conflicting performance of PZSIC for thermochemical properties 
    and barrier heights is called the {\em paradox of PZSIC}\cite{PERDEW20151}, resolution 
    of which was recently suggested by using the local scaling of the 
    exchange-correlation and Coulomb energy densities\cite{doi:10.1063/1.5129533}.
      A few schemes to rectify the over-correcting
      tendency of PZSIC have been proposed and examined.
      J\'onsson and coworkers\cite{doi:10.1063/1.4752229} scaled down 
      the entire SIC contribution by 50\%, and reported improved performance 
      in atomization energy. They also reported that using complex orbitals
      can improve the performance especially in case of LSDA. 
      In 2006, Vydrov \textit{et al.}\cite{doi:10.1063/1.2176608} proposed
      a method that scales down SIC in the many electron region using an
      iso-orbital indicator weighted by the density of local orbital. 
      To distinguish from the constant ({\em global}) scaling approach of J\'onsson
      and coworkers, we shall hereafter call the orbital dependent scaling approach by Vydrov and 
      coworkers as {\em orbital scaling} method.
      Vydrov and coworkers examined in detail the performance of various powers of 
      scaling factor for correcting the SIE in 
      the LSDA, Perdew, Burke, and Ernzerhof (PBE)\cite{PhysRevLett.77.3865}, Tao, Perdew,
      Staroverov and Scuseria (TPSS)\cite{PhysRevLett.91.146401},  and a hybrid of PBE with
      25\% of exact exchange (PBEh)\cite{doi:10.1063/1.478522, doi:10.1063/1.478401} functionals. 
      Subsequently, they also employed the orbital scaling to SIC to study the
      effect of scaled down SIC on the dissociation curves of H$_2^+$, 
      He$_2^+$, LiH$^+$, and Ne$_2^+$\cite{doi:10.1063/1.2566637}. They found that only the unscaled 
      PZSIC consistently yielded qualitatively correct curves for all four systems\cite{doi:10.1063/1.2566637}.
      Their orbital scaling approach to PZSIC is free from exact one- and nearly exact two-electron SI but 
      still suffers  many-electron SIE\cite{doi:10.1063/1.2566637}.
      Thus, benefit of the orbital scaling was primarily limited to 
      equilibrium properties. 
           
      Since the report of the work by Vydrov and coworkers, a number advances in the functional 
      development have been reported. One important advance at the meta-GGA level is the development 
     of {\em strongly constrained and appropriately normed} (SCAN) semilocal functional\cite{PhysRevLett.115.036402}.
      SCAN satisfies all 17 known exact constraints that a meta-GGA functional can satisfy.
      A number of studies reported in literature show that the SCAN functional provides improvement
      over other functionals for a wide variety of 
      solid-state and molecular properties\cite{PhysRevLett.121.207201,chen2017ab,PhysRevB.95.054111}.
      Recently, we investigated the performance 
      of the SCAN functional and self-interaction corrected SCAN functional for a wide array of molecular properties
      and found that eliminating self-interaction errors improves the performance of SCAN for dissociation 
      energies and barrier heights but it worsens the atomization energies\cite{doi:10.1063/1.5120532}. 
      The goal of the present work is multifold. We first want to examine the performance of orbital scaling when used with 
      SCAN meta-GGA functionals for various electronic properties such as total atomic energies, ionization potentials, electron affinities, molecules atomization energies, 
      reaction barrier heights, and dissociation and reaction energies. We also want to explore the use 
      of alternative scaling factors in order to see if they provide any improvement over the scaling factor 
      used by Vyrdov and coworkers. Finally, we want to explore if the orbital scaling approach can 
      be modified by differentially scaling the SIC for orbitals to obtain even better all-around performance.
      We illustrate this idea by proposing a
       new orbital scaling scheme that preserves correct $-1/r$  asymptotic behavior of the potentials for atoms. We also show that this new scaling scheme leads to significant improvements over the original orbital scaling approach  for number of properties.

\section{\label{sec:theory}Theory}
    The PZSIC method removes the SIE in the approximate density functionals by means of orbital-dependent corrections 
 to the approximate functional as follows, 
\begin{equation}\label{eq:PZSIC}
    E_{XC}^{PZSIC-DFA}=E_{XC}^{DFA}[\rho_\uparrow,\rho_\downarrow]-\sum_{i\sigma}^{occ}\left\{ U[\rho_{i\sigma}]+E_{XC}^{DFA}[\rho_{i\sigma},0] \right\}.
\end{equation}
Here, $\rho_{i\sigma}$ is the density of the $i^{th}$ orbital of spin  $\sigma$, and
$U[\rho_{i\sigma}]$ and 
$E_{XC}^{DFA}[\rho_{i\sigma},0]$ are the self-Coulomb and
the self-exchange-correlation energies.
In their 1981 work, Perdew-Zunger\cite{PhysRevB.23.5048} presented SIC calculations on the atoms using the orbital densities obtained from the KS orbitals. They also noted that delocalized nature of KS orbital for extended system will make the  SIC non size-extensive. Subsequently, Pederson, Heaton and Lin\cite{doi:10.1063/1.446959,doi:10.1063/1.448266} implemented PZSIC using local orbitals and performed the first SIC calculation on molecules. These localized orbitals are obtained from the unitary transformation of the KS orbitals by minimizing the energy which results in the Pederson 
{\em{localization}} equations:
\begin{equation}
             \langle \phi_{j\sigma}\vert V_{j\sigma}^{SIC} - V_{i\sigma}^{SIC}\vert
             \phi_{i\sigma}\rangle = 0 .
\end{equation}
         
 Fermi-L\"owdin orbital SIC (FLOSIC)\cite{doi:10.1063/1.4869581} is a recently proposed approach to remove the SIE using the PZSIC 
 methodology. In the FLOSIC, the optimal local orbitals, called Fermi-L\"owdin orbitals (FLOs),
 are obtained by a unitary transformation that depends on position-like variables such that 
 unitary invariance of the total energy is ensured.
 The Fermi orbitals are constructed  by introducing 
 the Fermi orbital  descriptor (FOD) positions\cite{Luken1982,Luken1984,Leonard1982}.
Using the FOD positions \textbf{a}$_j$, the KS orbitals $\psi_i$ are transformed into the Fermi orbitals 
 $\phi_j$ as follows, 
\begin{equation}
    \phi_j(\vec{r})=\frac{\sum_{\alpha}^{N} \psi_\alpha(\textbf{a}_j) \psi_\alpha(\vec{r})}{\sqrt{\rho(\textbf{a}_j)}} .
\end{equation}
Here, $N$ is the number of occupied orbitals. The localized Fermi orbitals $\{\phi_i\}$ are subsequently orthogonalized using
L\"owdin orthogonalization to obtain the FLOs.
By finding the optimal FOD positions that minimize the total energy, we can find the solution of Eq. (\ref{eq:PZSIC}).
The optimal positions of the FODs are 
obtained by minimizing the energy using either conjugate-gradient method or the L-BFGS algorithm\cite{Liu1989}.

   As mentioned in Sec. \ref{sec:introduction}, the application of PZSIC worsens the description
   of equilibrium properties when used with semilocal functionals. 
   To rectify  the overcorrecting tendency of PZSIC,
   Vydrov and coworkers\cite{doi:10.1063/1.2176608}
   scaled down the SIC in many-electron region using an  orbital dependent
   scaling factor,
$X_{i\sigma}^k=\int z_\sigma^k(\vec{r}) \rho_{i\sigma}(\vec{r})d\vec{r}$. 
Here, $k$ is an integer and  
$ z_\sigma(\vec{r})= {\tau_\sigma^W(\vec{r})}/{\tau_\sigma(\vec{r})},$
where 
$
    \tau_\sigma(\vec{r})=\frac{1}{2}\sum_i|\nabla\psi_{i\sigma}(\vec{r})|^2 ,
$
being the non-interacting kinetic energy density, 
and $\tau_\sigma^W$ is the von Weizs\"acker kinetic energy density.
The iso-orbital indicator  $z_\sigma$ is a function of position in space and 
interpolates between the uniform density region ($z_\sigma=0$)
and  one-electron region, $z_\sigma=1$. 
Vydrov \textit{et al.}\cite{doi:10.1063/1.2176608} recommend $k\geq3$ for the TPSS meta-GGA to preserve the correct fourth-order expansion in the limit of slow varying density.
The same consideration should apply to SCAN meta-GGA.
 This way of scaling down PZSIC with an orbital dependent scaling
 factor will referred hereafter as OSIC. 

The SIC energy in the OSIC approach of  Vydrov \textit{et al.}\cite{doi:10.1063/1.2176608} is given by
\begin{equation}\label{eq:orbsic}
    E^{OSIC}=-\sum_{i\sigma}^{occ}X_{i\sigma}^k \left( U[\rho_{i\sigma}] + E_{XC}^{DFA}[\rho_{i\sigma},0] \right).
\end{equation}
It is evident that OSIC reduces to PZSIC for  $k=0$.
In the $k\rightarrow \infty$ limit, Eq. (\ref{eq:orbsic}) becomes zero with an exception of one electron system. For the 
one-electron systems, the scaling factor will be 1 for any integer $k$. 
The  scaling factor $ z_\sigma(\vec{r})= {\tau_\sigma^W(\vec{r})}/{\tau_\sigma(\vec{r})},$
has the advantage that it vanishes in the uniform electron gas limit.
It is not the only choice for the scaling factor in OSIC. A number of alternative choices can be made.
Vydrov and Scuseria subsequently used the ratio of the orbital density to the total-spin density\cite{doi:10.1063/1.2204599}. This does not
require kinetic-energy densities and gave results comparable to those obtained using 
$ z_\sigma(\vec{r})= {\tau_\sigma^W(\vec{r})}/{\tau_\sigma(\vec{r})}$.
In this work, we also explore use of two other scaling factors. The first one is the electron 
localization function (ELF) introduced by Becke and Edgecombe\cite{becke1990simple}.
The ELF is commonly used for classifying chemical bonds and is defined as follows,
\begin{equation}
    \text{ELF} = \frac{1}{1+\alpha^2}
\end{equation}
where $\alpha=(\tau - \tau^W) / \tau^{unif}$,
and $\tau^{unif}=(3/10)(3\pi^2)^{2/3}\rho^{5/3}$ is $\tau$ in the uniform-density limit.
Using ELF in place of $z_\sigma$ partially satisfies the correct limits of the OSIC scaling factor. Although ELF = 1 for the single orbital limit, ELF = 0.5 in the uniform gas limit. 
Additionally, we also use $\beta$, another iso-orbital indicator defined as
\begin{equation}
    \beta = \frac{\tau - \tau^W}{\tau + \tau^{unif}},
\end{equation}
which has been used recently in construction of meta-GGA functionals\cite{PhysRevB.99.041119}.
Following how $\beta$ is used in functional design, we use $1-(2\beta)^2$ as the alternative
for $z_\sigma$ in $X_{i\sigma}^k$.  
Although this form can become negative, we included
it nonetheless for comparison since it has the correct interpolation between the single orbital
limit, $\lim_{\tau\to\tau^W} \{1-(2\beta)^2\} = 1$, and uniform-gas limit, $\lim_{|\nabla\rho|\to 0, \tau\to\tau^{unif}} \{1-(2\beta)^2 \}= 0$ since $\tau^W$ becomes $0$.
Thus, OSIC with this scaling factor also recovers the uniform gas limit as the SI-correction
vanishes in this limit and OSIC reduces to the DFA.

Appraisal of the orbital scaling factor for various electronic properties\cite{doi:10.1063/1.2176608}
showed that orbital scaling
requires different values of $k$ for different properties to obtain improved results. 
For example, excellent atomic energies are obtained for $k=4$, but $k=1$ or less is needed to obtain good estimates of reaction barrier heights.
The orbitalwise scaling down of PZSIC leads to violation of some
exact constraints satisfied in PZSIC. One such consequence is that it destroys the desirable correct 
$-1/r$ behavior of the exchange-correlation potential of the PZSIC. The orbital scaling
of Eq. (\ref{eq:orbsic}) also provides the poor performance for many-electron SIC\cite{doi:10.1063/1.2566637}
compared to the original PZSIC. As the asymptotic behavior is important in many physical process such as electron delocalization or in an accurate description of the charge transfer process, a new scaling approach
that preserves $-1/r$ asymptotic of the potential can be formulated. We refer to this approach as selective-scaling-OSIC (SOSIC). The SOSIC correction to the energy
in this approach is given by
\begin{equation}\label{eq:orbsicnew}
    E^{SOSIC}=-\sum_{i\sigma}^{M}X_{i\sigma}^k \left( U[\rho_{i\sigma}] + E_{XC}^{DFA}[\rho_{i\sigma},0] \right) \\
         -  \sum_{i\sigma}^{P} Y_{i\sigma} \left( U[\rho_{i\sigma}] + E_{XC}^{DFA}[\rho_{i\sigma},0] \right),
\end{equation}
  where $M=N-P$, with $N$ being the total number of occupied electrons. $P$ is the number local orbitals corresponding to the electrons in the 
   highest occupied orbital (HOO) shell.  For example, for $Ar$ atom $P=8$  as there are six electrons in
  the degenerate HOO shell that project onto 8 $sp^3$ local orbitals. We set $Y_{i\sigma}=1$ to maintain the accurate asymptotic description of
  exchange-correlation potential. We shall show later that this SOSIC
  essentially preserves the accuracy of unscaled PZSIC HOO eigenvalues and leads to overall improvements 
  of electronic properties in both the equilibrium cases as well as in stretched bond situations.
  The application of SOSIC requires identifying the FODs or FLOs that correspond to the HOO.
  This can be accomplished by finding the  FLO which has maximal overlap with the KS HOO. 
  We note that even though we used $Y_{i\sigma}$ to be unity its value can adjusted so that the negative of HOO eigenvalue
  matches with the exact experimental ionization potential. Adjusting the potential so that the magnitude of the HOO eigenvalue agrees
  with first ionization potential has been used  previously in the context of 
  fully analytic (grid free) Slater-Roothaan
  method\cite{zope2005slater}. In recent years, a similar procedure has been used to obtain the range separation parameters in many  applications of range separated hybrids
  method\cite{livshits2007well,baer2010tuned,kronik2012excitation,foster2012nonempirically,sun2013influence}.

  We implemented the OSIC and SOSIC method in the FLOSIC code\cite{FLOSICcode,FLOSICcodep} that is based on UTEP-NRLMOL. 
The Porezag-Pederson NRLMOL basis set\cite{PhysRevA.60.2840} which is roughly similar to quadruple zeta 
quality basis was used. FLOSIC uses a variational mesh\cite{PhysRevB.41.7453} that provides efficient numerical integration. 
The SCAN meta-GGA\cite{PhysRevLett.115.036402} was recently implemented\cite{doi:10.1063/1.5120532}
in the FLOSIC code.
We used very dense mesh tailored for SCAN that provides energy convergence with 
respect to the radial mesh within $10^{-8}$ Ha accuracy\cite{doi:10.1063/1.5120532}.
For calculations of anions, 
in addition to NRLMOL extra basis sets, 
long range s, p, and d single Gaussian orbitals
are used where their exponents ($\beta$) are extrapolated from $N^{th}$ basis of a given system using a relation as $\beta(N)^2/\beta(N-1)$.
This inclusion of additional diffuse exponents was suggested by Withanage \textit{et al.}  
\cite{PhysRevA.100.012505} for giving better descriptions of the extended nature of the anions. 
The full Hamiltonian in the OSIC is given by
\begin{eqnarray}\label{eq:OSICHAM}
    H_i = &-&\frac{1}{2}\nabla^2 + v(\vec{r})+\int \frac{\rho(\vec{r'})}{|\vec{r}-\vec{r'}|}d\vec{r'}+v_{XC}^{DFA}([\rho_\uparrow,\rho_\downarrow],\vec{r}) - X_i^k \left(
    \int \frac{\rho_i(\vec{r'})}{|\vec{r}-\vec{r'}|}d\vec{r'}+v_{XC}^{DFA}([\rho_i,0],\vec{r})
    \right) \nonumber\\
    & - &  z^{k}_\sigma(\vec{r}) \left(  U[\rho_i(\vec{r})] + E_{XC}^{DFA} [\rho_i(\vec{r}),0] \right)
    - \sum_{m}  \left(  U[\rho_m(\vec{r})] + E_{XC}^{DFA} [\rho_m(\vec{r}),0] \right) 
         \frac{\partial X_m^k}{\partial \rho(\vec{r}).}
\end{eqnarray}
In the present implementation, we ignore the last two terms which arise due to
variation of scaling factors in constructing the 
Hamiltonian. The calculations are performed self-consistently using Jacobi 
updates\cite{PhysRevA.95.052505} similar to
earlier reported FLOSIC calculations\cite{doi:10.1063/1.4996498,doi:10.1063/1.5120532,PhysRevA.100.012505,doi:10.1021/acs.jpca.8b09940,doi:10.1063/1.5125205,C9CP06106A,FLOSIC_WATER_PNAS,doi:10.1063/1.5050809,doi:10.1002/jcc.25767} but they are not full variational due to the
neglected terms.
We assessed the importance of the neglected terms by comparing our OSIC results with 
those of Vydrov and coworkers for the LSDA functional and obtained essentially the 
same results for various electronic  properties studied here. 
For instance, using the OSIC-LSDA with $k=1$, mean absolute error (MAE) per electron of total energies for 
Li--Ar is $0.004$ Ha in both methods, and MAEs for AE6 and BH6 are $18.0$ and $3.3$ kcal/mol with our 
implementation whereas Vydrov \textit{et al.} obtained $21.0$ and $3.5$ kcal/mol. The small differences 
can arise from the different choice of the basis sets used to obtained these MAEs.
These results also indicate that variations in the scaling factor are not too crucial for the properties studied here. 
Full variational calculation will be implemented in future.
Thus, the orbital SIC energies are scaled down as Eq. (\ref{eq:orbsic}) or Eq. (\ref{eq:orbsicnew}), and self-interaction correction to the Hamiltonian matrix elements of $i^{th}$ orbital are scaled down accordingly as Eq. (\ref{eq:OSICHAM}) by ignoring the last two terms.

The orbital scaling calculations performed this way has comparable computational 
cost as PZSIC. The only additional cost is the calculation of the scaling factor
which is not significant.
For SCAN calculations, the FODs used in this study were optimized at the FLOSIC-SCAN level of theory where a convergence tolerance of at least $10^{-6}$ Ha was used. 
For all orbital scaling calculations, the FOD positions and electron densities optimized at the FLOSIC-SCAN level of theory were used as a starting point for a given system. 
For the AE6, BH6, SIE4$\times$4, and SIE11 calculations, we used the geometries from the test sets as provided. The SIE4$\times$4 calculations required us to use simple mixing with a larger mixing parameter $\alpha=0.15-0.35$ to achieve SCF convergence.
OSIC-LSDA calculations were performed self-consistently in a similar fashion where we used FLOSIC-LSDA FODs and densities as a starting point.

\begin{figure}
    \centering
    (a)\includegraphics[width=0.9\columnwidth,trim = {0 0 0 0}, clip]{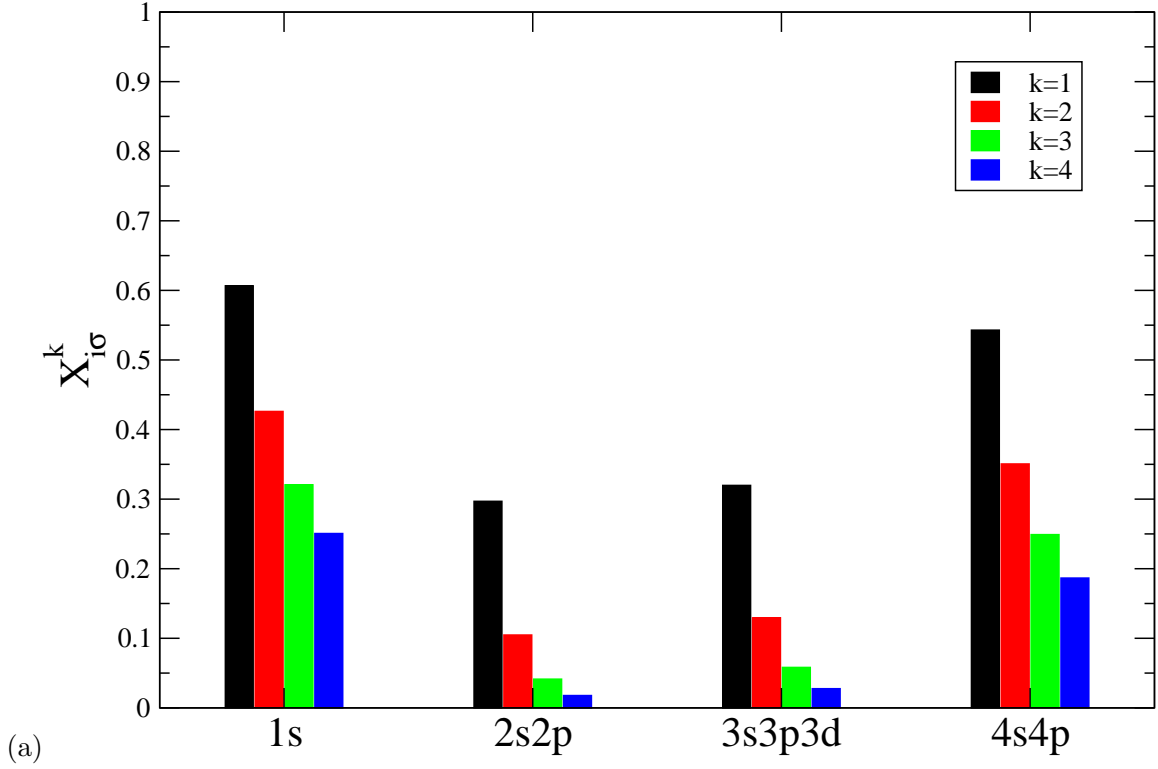}
    
    \vspace{1cm}
    
    (b)\includegraphics[width=0.9\columnwidth,trim = {0 0 0 0}, clip]{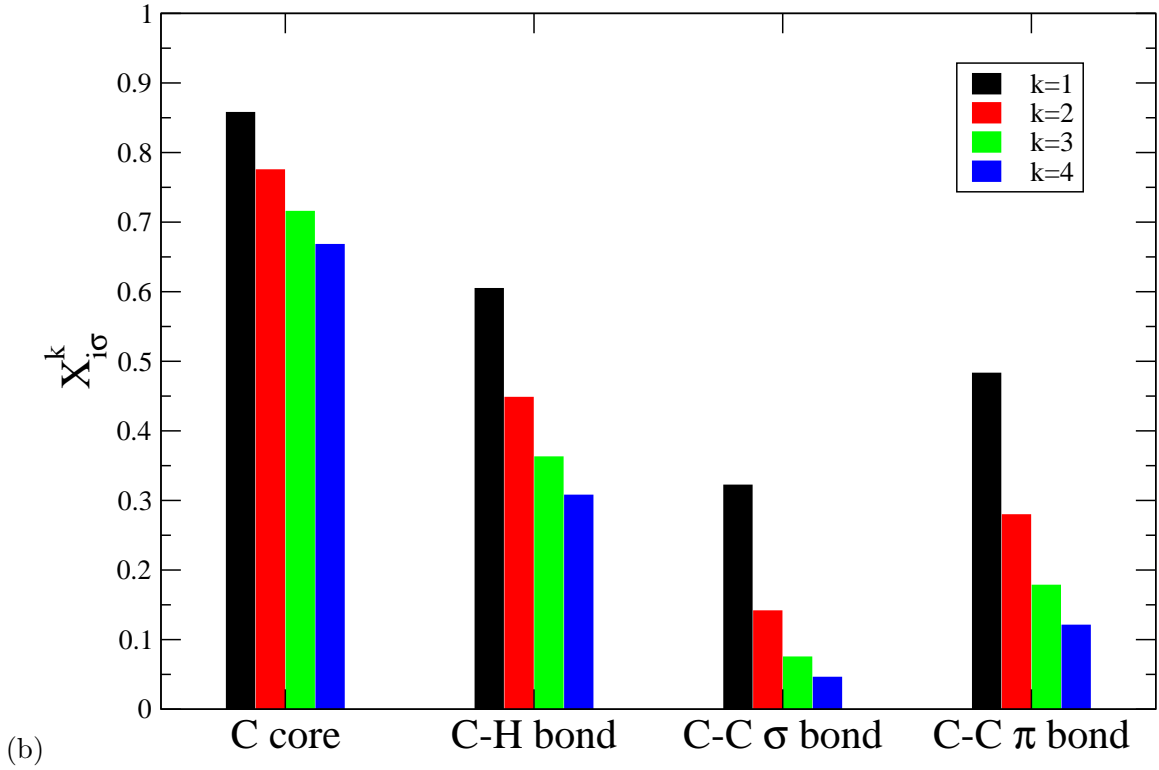}
    \caption{Scaling factors $X_{i\sigma}^k$ with $z_\sigma(\vec{r})={\tau_\sigma^W(\vec{r})}/{\tau_\sigma(\vec{r})}$ 
    and varying values of $k$: (a) The averaged values for each electron shells of Kr atom and (b) the average values for each bond type of benzene.}\label{fig:scalingfactor}
\end{figure}

\begin{figure}
\centering
\includegraphics[width=\columnwidth]{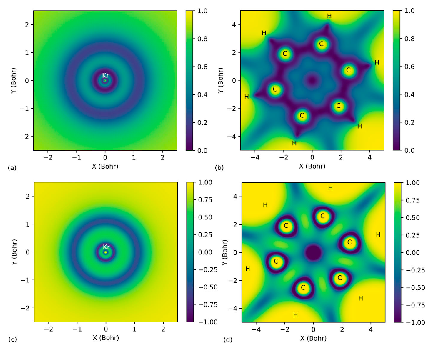}
  \caption{
  The contour map of $z_\sigma(\vec{r})={\tau_\sigma^W(\vec{r})}/{\tau_\sigma(\vec{r})}$ for (a) Kr atom and (b) benzene in OSIC-SCAN ($k=1$) calculations.
  Similarly, $1 - (2 \beta )^2$ is shown for (c) Kr atom and (d) benzene.
  $z_\sigma(\vec{r})=1$ for the single orbital regions, and $z_\sigma(\vec{r})=0$ for the uniform density regions. For simplicity, only the spin-up kinetic energy density ratio is shown.}
  \label{fig:ztau}
\end{figure}

Figure \ref{fig:scalingfactor} shows the values of the scaling factors $X_{i\sigma}^k$   
for Kr atom and benzene within OSIC-SCAN calculations.
In both cases, core orbitals tend to have a larger value than the rest. For the case of benzene, the factors for C-C $\sigma$ bonds and $\pi$ bonds have the values less than $0.5$.
The actual values of the OSIC scaling factor depend on two elements: (i) compactness of local orbital and (ii) size of the single orbital regions identified from the iso-orbital indicator. For instance, the scaling factor of core orbitals in benzene is larger than that of 1s orbital in Kr atom as can be seen from  
the contour maps of the scaling factors in Fig. \ref{fig:ztau}.

\section{\label{sec:results}Results}

\subsection{Atoms: total energies, ionization potentials, and electron affinities}
\begin{figure}
    \centering
    \includegraphics[width=\columnwidth]{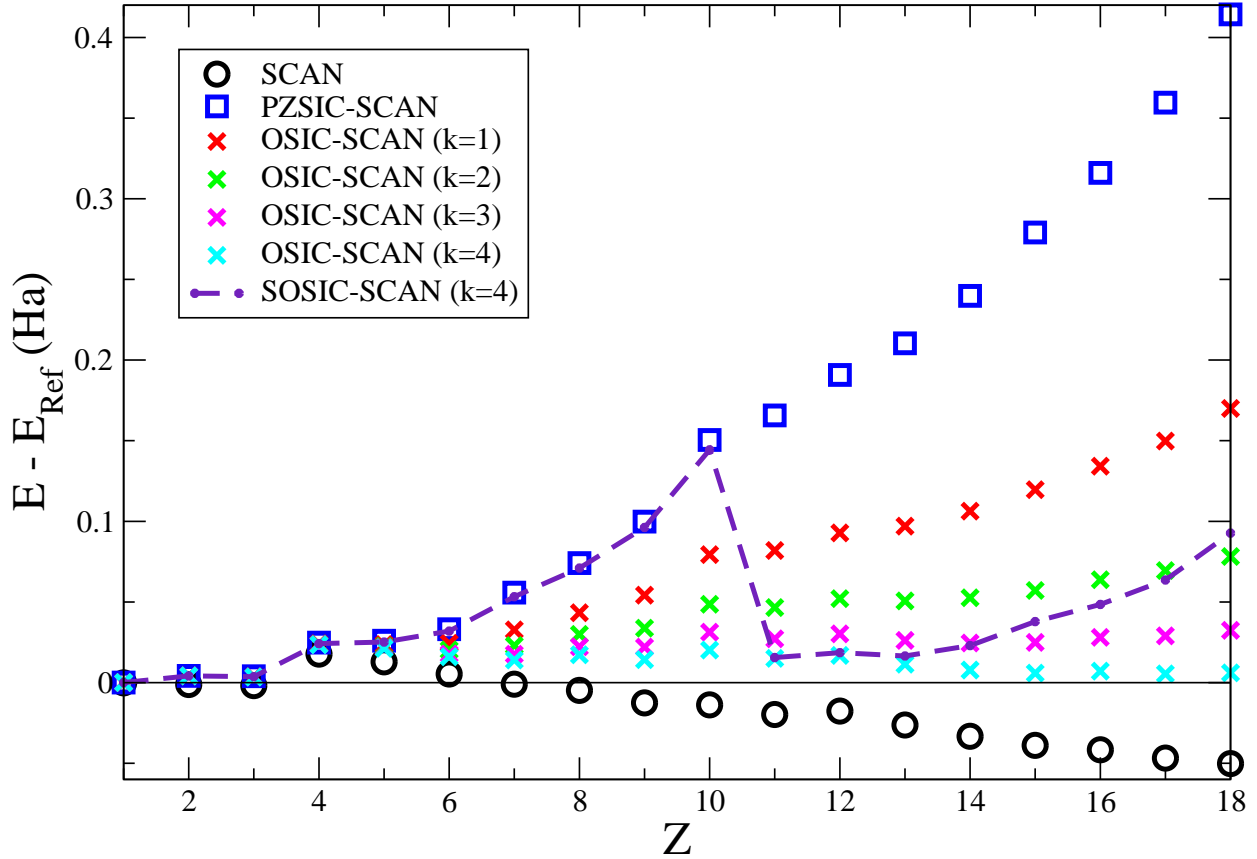}
    \caption{Total energy difference (in Hartree) of atoms Z$=1-18$ with respect to exact energies in various methods.}
    \label{fig:atoms}
\end{figure}

We studied atoms $Z=1-18$ and compared the total energies using different scaling powers on PZSIC and OSIC with accurate non relativistic calculated reference values obtained by Chakravorty \textit{et al.}\cite{PhysRevA.47.3649}.
The total energy differences with respect to reference values are shown in Fig. \ref{fig:atoms} for 
various values of $k$.
The PZSIC corresponds to $k=0$, and all $k\geq 1$ results shown are with the OSIC method. 
We include the recently reported DFA-SCAN and PZSIC-SCAN results   \cite{doi:10.1063/1.5120532} 
here for comparison.
DFA-SCAN shows small MAE of 0.019 Ha, and correcting
for self-interaction results in significantly over-corrected total energies (MAE, 0.147 Ha)
with systematic increase in error with increase in number of electrons. 
For Ar, the SI-correction is approximately 0.46 Ha. 
We note that application of OSIC reduces the MAE and as the $k$ values increases the MAE becomes smaller.
The results for different scaling powers are summarized in Table \ref{tabatoms}.
 OSIC-SCAN with $k=1$ reduces the MAE from 0.147 to 0.069 Ha,
whereas  $k=3$  shows comparable performance with DFA-SCAN.
The best total energies are obtained with $k=4$ with MAE of only 0.012 Ha.
These results are consistent with earlier reports\cite{doi:10.1063/1.2176608} that increasing value of $k$ results in 
better total energies.
A comparison with previously reported results show that OSIC-SCAN with $k=4$
gives atomic energies that are better than the previously reported OSIC-LSDA, OSIC-PBE, or OSIC-TPSS.
The MAE per electron of atoms Li to Ar for OSIC-SCAN $k=4$ is 0.002 Ha whereas 
the reported best MAEs for OSIC-LSDA $k=1$, OSIC-PBE $k=3$, 
and OSIC-TPSS $k=3$ are 0.004, 0.007, and 0.003 Ha respectively.

\paragraph{Ionization Potentials}
Ionization potentials (IP's) and electron affinities (EA's) are determined by processes that 
involve electron removal and electron addition, respectively. These 
processes are therefore sensitive to the asymptotic structure of the effective
potential. One would therefore expect that removal of self-interaction 
in DFAs would result in significant improvement in the quality of 
these quantities. In general, however it has been found that these 
quantities calculated as the difference of total energies of neutral 
and charge system are fairly accurately predicted by many DFAs.
The calculation of IPs (EAs) from the self-consistent 
total energy differences of  atoms and their cation (anion) is called $\Delta$SCF method\cite{PhysRevB.13.4274}.
We calculated the IPs for atoms from Hydrogen through Krypton using the $\Delta$SCF method and compared them
against the experimentally reported values from Ref. [\onlinecite{NIST_CCCBD}] in Table \ref{tabip}.
To facilitate a direct comparison with the values reported by Vydrov \textit{et al.}, we also present results for a subset of atoms, from Hydrogen
through Argon.
For this smaller subset of atoms $Z=2-18$, the MAEs are 0.175 and 0.274 eV for SCAN and PZSIC-SCAN respectively. 
The OSIC-SCAN results for various $k$ values have similar performance with MAEs within $0.178-0.181$ eV. 
Vydrov \textit{et al.}\cite{doi:10.1063/1.2176608} reported that the OSIC with $k=2$ and $3$ improves IPs of atoms $Z=1-18$ for 
LSDA, PBE, TPSS, and PBEh functionals. A similar trend was also observed in the present
OSIC-SCAN ionization potentials for the subset of atoms.
For this subset, the DFA already performs well and OSIC lowers the larger errors produced by PZSIC 
bringing the resultant errors close to those in DFA and in some cases improves them further.
However, if one extends the data set to include a larger number of atoms ($Z=2-36$),
then a different trend is observed. In this case,
PZSIC-SCAN (MAE $~0.259$ eV) shows better performance than DFA-SCAN (MAE $~0.273$ eV).
The OSIC-SCAN results have 
MAEs ranging $0.304-0.349$ eV, and these errors are larger than both DFA and PZSIC. 
This result suggests that full SIC treatment is needed  to obtain accurate 
estimates of IPs of heavier atoms.
All OSIC results for the complete set ($Z=2-36$) studied here have similar errors as we have seen for the smaller subset ($Z=2-18$), but there is a slight but noticeable decrease in errors as the value of $k$ increases.

\paragraph{Electron Affinity}
We studied EAs of 20 atoms, specifically atoms, H, Li, B, C, O, F, Na, Al, Si, P, S, Cl, K, Ti, Cu, Ga, Ge, As, Se, and Br. These 20 atoms are experimentally shown to bind an extra electron, and their experimental EAs are found in the NIST database in Ref. [\onlinecite{NIST_CCCBD}].
Similar to the IP calculations, EAs were obtained using the $\Delta$SCF approach. 

\begin{figure}
    \centering
    \includegraphics[width=1.0\columnwidth]{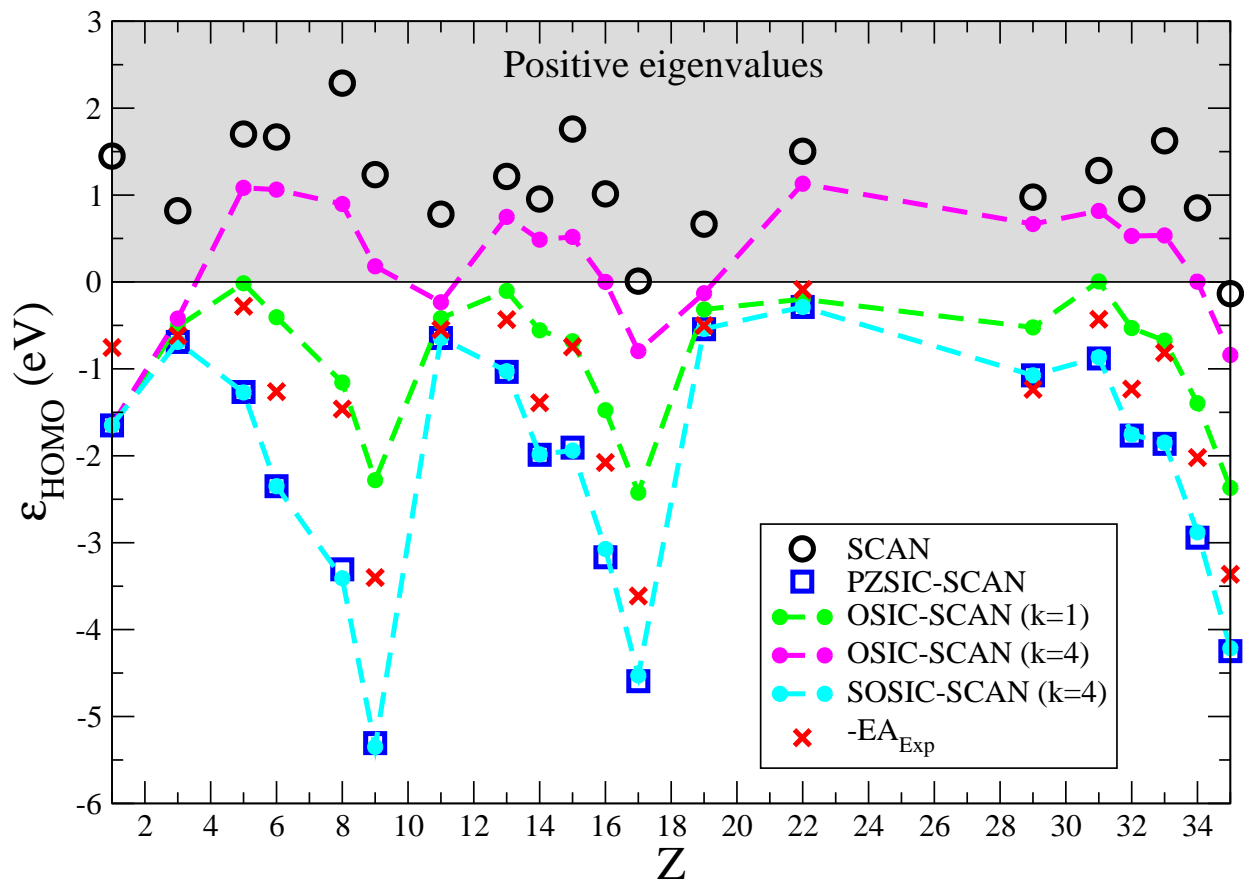}
    \caption{The HOO eigenvalue of 20 atoms within various models along with 
    negative of experimental EA values (in eV).}
    \label{fig:anionhomo}
\end{figure}

In Table \ref{tabea}, we present results for a subset of 12 EAs for the first three rows of periodic table and the third column shows the results  
that include the fourth row in addition to the 12 EAs resulting in 20 EAs.
For the 12 EAs, DFA-SCAN shows the smallest error but it has the problem of positive HOO eigenvalues. 
Correcting for SIE results in binding of the electron, and PZSIC-SCAN shows the MAE of 0.364 eV for $\Delta$SCF EAs. The OSIC-SCAN with $k=4$ improves the EAs to MAE of 0.125 eV.
For the larger set of 20 EAs, MAEs are 0.148 and 0.341 eV for SCAN and PZSIC-SCAN in the respective order. 
The OSIC-SCAN gives performance improvement especially when $k=2$ is used. The error in this case is the smallest with MAE of 0.128 eV. Thus, OSIC-SCAN provides better performance for the EAs than the PZSIC-SCAN.
We note that although the $\Delta$SCF approach yields positive EAs for the DFAs, the eigenvalue 
corresponding to the added electron becomes positive in all DFA anion calculations, indicating that the 
extra electron is not actually bound in the complete basis set limit.  This problem is due to the
incorrect asymptotic form of the potential in the DFA calculations.  SIC fixes this \cite{PhysRevB.23.5048}, leading to bound states for the HOO in the anions. As mentioned 
in the introduction, the OSIC  has undesirable effect on the asymptotic potential.
In OSIC, the correct $-1/r$ behavior of asymptotic potential in PZSIC 
is replaced by  $-X_{HO}/r$ where $X_{HO}$ 
is the scaling factor for the electronic shell to which HOO belongs.
In Fig. \ref{fig:anionhomo}, we compared the HOO eigenvalues for PZSIC and OSIC calculations 
along with the experimental electron affinity.  PZSIC gives negative HOO eigenvalues for all systems indicating the HOO electrons are bound to those atoms. It is evident from the figure that 
the absolute HOO eigenvalue in PZSIC overestimates the electron affinity. Applying OSIC shifts 
the eigenvalues upward. This upward shift for $k=1$ reduces the overestimation of absolute 
HOO as seen in PZSIC and bring it closer to the experimental electron affinities. But the 
shift systemically increases with the scaling factor as $k$ increases. As a consequence, the sign of the eigenvalue eventually
changes for some systems and electron in HOO becomes unbound as the asymptotic
potential becomes too shallow to provide sufficient attractive potential for 
the electron. This behavior was not noted earlier in the OSIC calculations of Vydrov and coworkers
but it was expected as scaling down SIC by larger factors brings OSIC results closer to those 
of DFAs. The OSIC with $k=4$ has a drawback that several atomic anions are unbound in this model.
Exceptions are  alkali metals and halogens who maintained negative eigenvalues with OSIC unless
very large scaling  power $k$ is applied. These exceptions occur as halogens
have larger electron affinities and because the scaled down 
factor for the HOO of alkali anions are large (e.g.,  0.83 for $Li^-$
and 0.71 for Na$^-$)  even for $k=4$.
For the rest of the atom families, anion HOO eigenvalues become positive with a scaling power of 2--3. This is not too surprising considering OSIC recovers DFA performance in the $k\rightarrow\infty$ limit. 

\subsection{Atomization energies}
The AE6 benchmark set\cite{doi:10.1021/jp035287b} was used to study the performance of the OSIC approach in atomization energies.
This set includes six molecules: SiH$_4$, SiO, S$_2$, propyne (C$_3$H$_4$), glyoxal (C$_2$H$_2$O$_2$), and cyclobutane (C$_4$H$_8$), and it is a good representation of the larger main group atomization energy (MGAE109) set\cite{doi:10.1063/1.3663871}. The geometries and reference values are obtained at the QCISD/MG3 level of theory. The atomization energy ($AE$) of a molecule is obtained with
\begin{equation}
    AE = \sum_i^{N} E_i - E_{mol} > 0
\end{equation}
where $E_i$ is the energy of atom, $E_{mol}$ is the energy of a given molecule, and $N$ is the number of atoms in the molecule.
$AE$s were compared against non spin-orbit coupling reference values reported in Ref. [\onlinecite{doi:10.1063/1.3663871}]. 
The results are summarized in Table \ref{tabAE6}.
The DFA-SCAN provides quite accurate estimates of $AE$ with MAE of only 2.85 kcal/mol. 
However, correcting for SIE worsens $AE$s. 
PZSIC-LSDA FODs were used for the MAE reported in Ref. [\onlinecite{doi:10.1063/1.5129533}]. It is found that 
relaxation of FODs within PZSIC-SCAN increases the error 
 (MAE ~ 26.52 kcal/mol) using the data extracted from Ref. [\onlinecite{doi:10.1063/1.5120532}]. We observed that scaling down PZSIC improves the performance as the value of $k$ increases with $k=4$ yielding the best MAE of $4.10$ kcal/mol; although this error is larger in comparison than that of DFA-SCAN, the value is six times smaller compared to the PZSIC-SCAN result. 
PZSIC-SCAN tends to overestimate total energies especially for molecules,
and this leads to a large discrepancy in atomization energies. Scaling down PZSIC helps reducing the overestimation and improves predicting atomization energies. This result shows that DFA-SCAN is better for predicting atomization energies without the self-interaction correction. Atomization energy calculations involve equilibrium molecular structure where the SCAN functional performs well.  

\subsection{Reaction barrier heights}\label{sec:bh6}

BH6 benchmark set \cite{doi:10.1021/jp035287b} was used to study the scaling down performance in reaction barrier. The BH6 set consists of three hydrogen transfer reactions (OH + CH$_4 \rightarrow$ CH$_3$ + H$_2$O, H + OH $\rightarrow$ O + H$_2$, and H + H$_2$S $\rightarrow$ H$_2$ + HS).  
Total energies for the left- and right-hand side and saddle-point of a given reaction formula were calculated, and the barrier heights of forward (f) and reverse (r) reactions were obtained from the energy differences of these three points. Errors are summarized in Table \ref{tab:errBH6}.

Many DFA functionals including SCAN do not give a correct picture of chemical reaction because in most cases the saddles points energies are underestimated. These are the cases where self-interaction correction becomes important. From the table it can be seen that 
PZSIC corrects the shortcoming of DFA in this situation. The full SIC treatment with $k=0$ reduces both ME and MAE. 
When orbital scaling is applied to PZSIC i.e., OSIC with $k=1$, the MAE  increases to 3.96 kcal/mol
from the PZSIC's 2.96 kcal/mol.  
The  MAE  systematically increases with higher powers of the scaling factor. 
In all cases of OSIC, the reaction barriers are underestimated for all six reactions
as can be seen from the MEs and MAEs in Table \ref{tab:errBH6}. 
For the saddle-point calculations with stretched bonds, one needs full SIC correction. The increase in value of $k$ results in larger 
percent of  SIC correction being scaled down which leads to poor estimates of barrier heights. Note that for $k=4$, the 
error is comparable to the DFA error.
A further discussion is presented in Sec. \ref{sec:discussion}.

\subsection{SIE benchmark sets}
The SIE11 sets consist of five cationic and six neutral chemical reactions that are very sensitive to self-interaction errors\cite{doi:10.1021/ct900489g}.
The SIE4$\times$4 sets consist of dissociation energy calculations of positively charged dimers at four different distances $R$ from their equilibrium distances $R_e$: $R/R_e =  1.0, 1.25, 1.5$, and $1.75$\cite{C7CP04913G}. 
Reaction energies for SIE11 and dissociation energies for SIE4$\times$4 were computed and compared against the reference values.
The dissociation energies and reaction energies are obtained from energy difference between left- and right-hand sides of a given chemical reaction formula.
The reference values provided in Ref. [\onlinecite{doi:10.1021/ct900489g}]  obtained at the coupled-cluster single double and perturbative triple [CCSD(T)]/CBS level of theory are used for comparison with our values.
The results are presented in Table \ref{tab:dissociation}.

From SCAN to PZSIC-SCAN, there is substantial decrease in errors: 
for SIE4$\times$4, the MAE is decreased from 17.9 to 2.2 kcal/mol.
Similar performance improvements are also seen in the SIE11 test set where 
the MAEs decrease from 10.4 to 5.1 kcal/mol for the cationic reactions and 
from 9.9 to 6.2 kcal/mol for the neutral reactions. 
On the other hand,
all of the OSIC results show larger errors compared to PZSIC-SCAN. 
Especially, for SIE4$\times$4 and SIE11 cationic reactions, larger MAEs are seen for higher $k$. 
For the SIE11 neutral systems, however, the error decreases for larger values of $k$ though it is still larger than the PZSIC-SCAN.

In our previous study\cite{doi:10.1063/1.5129533}, we used a pointwise local scaling approach on PZSIC-LSDA for the SIE sets. We found MAEs of 2.6, 2.31, and 6.31 kcal/mol for SIE4$\times$4, SIE11 cationic, and SIE11 neutral reactions respectively. In all three cases, deviations were decreased from PZSIC-LSDA.
In contrast, our OSIC results in Table \ref{tab:dissociation} show increase in errors going from PZSIC-SCAN to OSIC-SCAN.
We find that orbital scaling does not perform well for the SIE4$\times$4 and SIE11 calculations, 
while LSIC\cite{doi:10.1063/1.5129533}, which is an interior scaling approach, does not experience the same performance degradation.
The ideas to improve upon these shortcomings are discussed in Sec. \ref{sec:discussion}.

\section{Performance of different scaling factors}
   In Table \ref{tabatoms}, we compare the results of OSIC-LSDA with different iso-orbital indicators. For this comparison, we used $k=1$. 
We used ELF and $1-(2\beta)^2$ as alternatives for the scaling factor in OSIC. 
We investigated the effect of these scaling factors in OSIC-LSDA and OSIC-SCAN 
on four different properties: total energies of atoms, atomization energies (AE6), barrier heights (BH6), and SIE sets of reactions. 
With OSIC-LSDA, $1-(2\beta)^2$ produces larger MAE of 0.062 Ha in the total energies of atoms compared to ELF (0.037 Ha) and $\tau^W/\tau$ (0.035 Ha).
However, for the other properties, $1-(2\beta)^2$ shows better performance than the others. For atomization energies, the factor $1-(2\beta)^2$ yields an  MAE of 11.7 kcal/mol compared to MAE of  23.2 and 18.9 kcal/mol for ELF and  $\tau^W/\tau$, respectively. 
Similarly for barrier heights, $1-(2\beta)^2$ (MAE, 2.3 kcal/mol) shows better performance than ELF (MAE, 3.2 kcal/mol) and $\tau^W/\tau$ (MAE, 3.3 kcal/mol).
A large difference can be seen for SIE11 where MAE is 5.9 kcal/mol for $1-(2\beta)^2$.
This is almost half of PZSIC-LSDA MAE of 11.7 kcal/mol whereas the other two scaling factors show larger error than PZSIC-LSDA.

In addition to OSIC-LSDA, we also studied the effect of alternative scaling factors with OSIC-SCAN.
For OSIC-SCAN, all three scaling factors have comparable performance in atomic total energies, AE6, and BH6. There are some differences for the SIE sets where ELF is similar to $\tau^W/\tau$ ($k=3$) and $1-(2\beta)^2$ is similar to $\tau^W/\tau$ ($k=2$) in performance. Overall, the performance of PZSIC is best for the SIE sets of reactions.

\section{\label{sec:discussion} Discussions and improvements to the OSIC}

          Applications of OSIC in the present work to the SCAN functional show that the OSIC 
can overcome the worsening effects of the PZSIC results for equilibrium properties such as atomization
energies or total energies if higher values of $k$ are used. For example, OSIC-SCAN with 
$k=4$ gives good total energies and atomization energies. On the other hand, The OSIC-SCAN with same $k=4$, results in deterioration
of barrier heights or dissociation energies where the
bonds are stretched. In this case,
unscaled PZSIC (OSIC with $k=0$) works better than all scaled down PZSIC studied 
herein. Thus, no single value of $k$ is sufficient to obtain good results for 
all properties. These results are consistent with earlier scaled down PZSIC 
calculations of Vydrov and coworkers. The explanation as to why PZSIC does not perform well
for semi-local functionals has been understood in terms of the orbital densities.  It was shown that noded orbital densities
produce large errors when used to estimate the self-interaction correction using Perdew-Zunger
method\cite{doi:10.1063/1.2176608,doi:10.1063/1.5087065}.
It was found that these errors can be reduced but not eliminated 
using nodeless densities of complex orbitals. Another source of error in PZSIC
is that its application  to a semilocal functional causes appropriate
norms that are built in to the functional to be violated\cite{doi:10.1063/1.5090534}.
With OSIC, the loss of uniform electron gas limit depends on the form of the scaling factor used 
to identify many-electron region.
As discussed in Sec. \ref{sec:theory}, except for the ELF scaling factor used in this
work, the OSIC has the correct uniform electron gas limit. 
The OSIC approach shows behavior that is opposite to the paradoxical 
behavior of original PZSIC. It improves some properties (equilibrium properties) at the cost of worsening the barrier heights where the bonds are stretched. Recent interior local scaling LSIC approach which corrects for the self-interaction 
in single-orbital region by scaling energy densities does not suffer from such 
conflicting behavior\cite{doi:10.1063/1.5129533}.
The OSIC thus has limited usefulness over PZSIC unless property dependent 
choice of $k$ (powers of scaling factor) is made. The external scaling form of OSIC  (Eq. (\ref{eq:orbsic}))
offers  unique ways to apply the SIC (Eq. (\ref{eq:orbsicnew})).
For example, the paradoxical behavior of OSIC can be mitigated by 
selectively applying the orbital scaling factor used in each local orbital. That is, one can apply the 
scaling with $k=4$ for most orbitals (core and part of valence states) and keep the full 
PZSIC correction for the orbitals that  require full SIC treatment. We considered a few cases to demonstrate the potential of this approach which are discussed below.

\subsection{HOO eigenvalues}
\begin{figure}
    \centering
    \includegraphics[width=1.0\columnwidth]{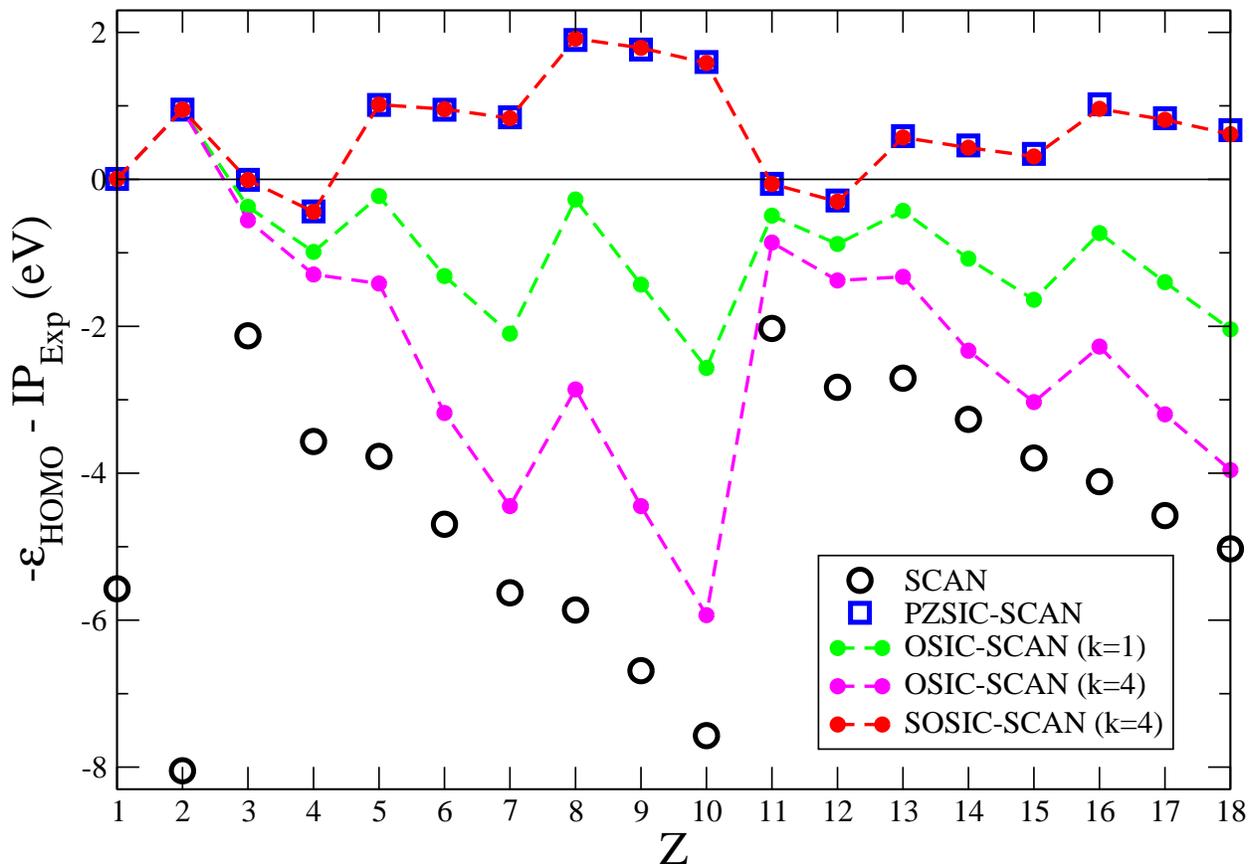}
    \caption{The difference in HOO eigenvalue of atoms $Z=1-18$ with respect to experimental IP (in eV). Note that,  unlike OSIC eigenvalues,
     SOSIC eigenvalues closely mimic PZSIC eigenvalues.}
    \label{fig:homocompare}
\end{figure}

One known shortcoming of orbital scaling is that the magnitude of the highest occupied eigenvalues ($\epsilon_{HO}$) becomes underestimated. 
In exact DFT, the highest occupied eigenvalue equals the negative of the ionization 
potential\cite{perdew1982density,levy1984exact,almbladh1985exact,perdew1997comment,PhysRevB.60.4545}. This relationship 
does not strictly hold for approximate density functionals and in most DFAs, the absolute value of the 
HOO eigenvalue substantially underestimates the first ionization potential due to SIE. 
In Table \ref{tab:ehomo}, we compared MAEs of the HOO eigenvalues of atoms $Z=1-36$ against the experimental IPs\cite{NIST_CCCBD} using several different methods. PZSIC shows the smallest MAE of 0.606 eV as expected from the PZSIC's correct asymptotic potential shape, 
and the OSIC (scaled down PZSIC with $k>0$) generally shows larger deviations.  
This arises due to the scaling down the correction for the highest occupied orbital. The correct asymptotic behavior 
can be preserved if Eq. (\ref{eq:orbsicnew}) is used instead of Eq. (\ref{eq:orbsic}).
To illustrate this, we applied the orbital scaling to PZSIC except for local orbitals on the electron shell that belong to the outermost electrons.  
The full PZSIC is used for these outermost orbitals. 
A comparison of the HOO eigenvalues of atoms so obtained are compared against experimental IPs\cite{NIST_CCCBD} for a smaller subset of atoms with $Z=1-18$ is presented in Fig. \ref{fig:homocompare}.
For this set, the OSIC with $k=4$ has an MAE of 2.414 eV which is significantly
larger compared to the PZSIC (MAE $~$ 0.763 eV).
On the other hand, SOSIC has MAE of only 0.754 eV which shows that  
SOSIC can provide the $-\epsilon_{HO}$ of same quality as the PZSIC. 
It is interesting to 
see how the SOSIC affects total energies. We have shown this for atoms in Fig. \ref{fig:atoms} (SOSIC-SCAN ($k=4$)). Since orbitals other than those belong to HOO 
shell are scaled, the total energy in SOSIC would lie between OSIC $k=4$ and PZSIC 
total energies. Thus, lighter atoms for which most of the orbitals belong to the HOO
shell have total energies closer to PZSIC.
For benefiting both accuracy of PZSIC's $-\epsilon_{HO}$ and improved total energies
from the orbital scaling, the best case is when a small fraction of local orbitals is mapped to HOO and is treated with full PZSIC. This is the case for the alkali metal atoms.
In the worst case, SOSIC recovers the PZSIC energies. Halogens and noble gases 
atoms are the examples of such case (See Fig. \ref{fig:atoms}).

\subsection{Barrier height}
 Barrier height is another property that can benefit from selective scaling in OSIC. From Sec. \ref{sec:results}, we know that OSIC
 with $k=4$ give good atomization energies but poor barrier heights. The  calculations of saddle points with stretched bonds 
 are responsible for the increased discrepancies in the BH6 benchmark result.
To see if barrier height estimates 
can be improved using selective scaling, we calculated the BH6 barrier heights using the following 
approach. Since we know that exterior scaling works well for the local orbitals in an equilibrium state, we applied the scaled down PZSIC ($k=4$) for these orbitals while using  full PZSIC for the orbital corresponding to the hydrogen transfer.
With this selective scaling, we obtained MAE of $1.92$ kcal/mol and ME of $-0.75$ kcal/mol. Curiously, this error is even smaller than MAE of  $2.96$ kcal/mol with PZSIC. This finding suggests that 
good results for barrier height calculation using the OSIC method can be achieved 
if scaling factors for
certain orbitals are chosen according to the characteristics of orbitals as the spirit of SOSIC.

 Finally, we comment on the possible effect of SOSIC on the dissociation energy curves. As 
 mentioned in the introduction,  Ruzsinszky and coworkers have studied the dissociation energy 
 curves of H$_{2}^{+}$, He$_2^+$, LiH$^+$, and Ne$_2^+$ using OSIC and have noted that unlike PZSIC, OSIC
 does not provide qualitatively correct curves for all four systems\cite{doi:10.1063/1.2566637}.
 The SOSIC may correct this failure of OSIC as it provides correct asymptotic description 
 of the potential. Our attempts to compute the dissociation curves for 
 LiH$^+$, and Ne$_2^+$ were not successful due to difficulties in obtaining 
 convergence in far stretched regime using the Jacobi scheme of Ref. [\onlinecite{PhysRevA.95.052505}]. A new method with a single SIC Hamiltonian is being developed\cite{SingHam} which shows promise in handling dissociating fractions with correct charge. We will study dissociation with SOSIC in future.

\section{\label{sec:conclusions}Conclusions}
   We have implemented the orbitalwise scaling down of PZSIC 
using the FLOSIC methodology.  
The OSIC method
is used in combination with SCAN meta-GGA functional to assess its performance
for a wide array of properties---for atoms: total energies, ionization potential, 
electron affinities, and for molecules: atomization energies, reaction barrier heights,
and dissociation energies. We find that for equilibrium properties the OSIC with 
$k=4$ works well, and it recovers the performance of the uncorrected SCAN. For 
non-equilibrium properties, we observed that full PZSIC treatment is necessary 
in many situations.  
The comparison of present 
OSIC-SCAN results with earlier reported OSIC-PBE and OSIC-TPSS meta-GGA\cite{doi:10.1063/1.2176608}
indicate superior performance of OSIC-SCAN over the OSIC-PBE and OSIC-TPSS.
We also show that by selectively scaling down and applying full PZSIC 
correction on active or outermost 
orbitals, the inconsistencies of  OSIC can be mitigated or
eliminated and its performance can be improved beyond equilibrium properties.
Thus, selective scaling down approach presented here can provide good description
of equilibrium properties, estimates of ionization energies from the HOO eigenvalues, 
stable atomic anions, and reaction barrier heights.  
The SOSIC thus 
provides major improvement over the OSIC formalism. It is interesting to 
compare the SOSIC approach with the LSIC method that we recently proposed\cite{doi:10.1063/1.5129533}.
The LSIC method removes the self-interaction selectively in spatial region 
where the correction is necessary and resolves 
the paradoxical behavior of PZSIC. It provides good results
for both equilibrium properties as well as for properties where bonds are stretched.
The SOSIC approach, though not as elegant as LSIC\cite{doi:10.1063/1.5129533}, accomplishes this goal
by choosing the scaling factors according to the characteristic of orbitals.
We hope that the present results along with our recent 
results\cite{doi:10.1063/1.5129533,doi:10.1063/1.5120532} 
provide more sanguine
future of SIC-DFA that has broader applicability than the standard DFAs.

\begin{acknowledgments}
Authors acknowledge Profs. Koblar Jackson, Adrienn Ruzsinszky, 
Mark Pederson and John Perdew for comments on the manuscript.
Discussions with    Mr. Carlos Diaz and  Dr. Luis Basurto 
are gratefully acknowledged.
R. R. Z. is grateful to Prof. Rajeev K. Pathak for introducing him to self-interaction-corrected density functional theory. 
This work was supported by the US Department of Energy, Office of 
Science, Office of Basic Energy Sciences, as part of the 
Computational Chemical Sciences Program under Award No. 
DE-SC0018331.
Support for computational time at the Texas Advanced 
Computing Center through NSF Grant No. TG-DMR090071, 
and at NERSC is gratefully acknowledged.
\end{acknowledgments}

\section*{Data Availability Statement}
The data that support the findings of this study are available from the corresponding author upon reasonable request.

\newpage
\section*{\label{sec:tables}Tables}
\begin{table}[!htb]
\caption{\label{tabatoms}
The mean absolute error (MAE) of total atomic energies in various methods. These MAEs are in Hartree atomic unit.
}
\footnotetext{Reference [\onlinecite{doi:10.1063/1.2176608}].}
\footnotetext{MAE reported in the reference is for atom $Z=3-18$.}
\footnotetext{Reference [\onlinecite{doi:10.1063/1.5129533}].}
\begin{ruledtabular}
\begin{tabular}{lcccc}
\textrm{Method}&
\textrm{LSDA}&
\textrm{SCAN}&
\textrm{PBE$^\text{a,b}$}&
\textrm{TPSS$^\text{a,b}$}\\
\colrule
DFA          & 0.726 (0.822)$^\text{a,b}$   & 0.019 & 0.101 & 0.022 \\ 
PZSIC        & 0.380 (0.420)$^\text{a,b}$   & 0.147 & 0.183 & 0.278 \\
OSIC ($k=1$) & 0.035 (0.037)$^\text{a,b}$  & 0.069 & 0.118 & 0.131 \\
OSIC ($k=2$) & (0.205)$^\text{a,b}$   & 0.038 & 0.095 & 0.073 \\
OSIC ($k=3$) & (0.316)$^\text{a,b}$   & 0.021 & 0.085 & 0.042 \\
OSIC ($k=4$) &                      & 0.012 &       & \\ 
\colrule
SOSIC ($k=4$) &   & 0.043 &   & \\ 
\colrule
OSIC ($z_\sigma =$ ELF, $k=1$)          &  0.037  & 0.069 & &\\
OSIC [$z_\sigma = 1-(2\beta)^2$, $k=1$] &  0.062  & 0.063 & & \\
\colrule
LSIC & 0.041$^\text{c}$ & & & \\
\end{tabular}
\end{ruledtabular}
\end{table}

\begin{table}[!htb]
\caption{\label{tabip}
The mean absolute error (in eV) of $\Delta$SCF ionization potentials computed in various methods.
}
\footnotetext{Reference [\onlinecite{doi:10.1063/1.2176608}].}
\footnotetext{Reference [\onlinecite{doi:10.1063/1.5129533}].}
\begin{ruledtabular}
\begin{tabular}{lcccc}
&\multicolumn{2}{c}{SCAN}& PBE$^\text{a}$ & TPSS$^\text{a}$ \\ \cmidrule(lr){2-3} \cmidrule(lr){4-4} \cmidrule(lr){5-5}
\textrm{Method}&
\textrm{Z=2--18 (17 IPs)} &
\textrm{Z=2--36 (35 IPs)} &
\textrm{Z=1--18 (18 IPs)} &
\textrm{Z=1--18 (18 IPs)} \\
\colrule
DFA &0.175& 0.273 & 0.15 & 0.12 \\
PZSIC        &0.274& 0.259 & 0.39 & 0.34\\
OSIC ($k=1$) & 0.181 & 0.342 & 0.22 & 0.17\\                        
OSIC ($k=2$) & 0.178 & 0.322 & 0.15 & 0.12\\
OSIC ($k=3$) & 0.178 & 0.304 & 0.12 & 0.11\\                        
OSIC ($k=4$) & 0.183 & 0.294 & &\\
\colrule
&\multicolumn{2}{c}{LSDA}&  &  \\ \cmidrule(lr){2-3} 
\textrm{Method}&
\textrm{Z=2--18 (17 IPs)} &
\textrm{Z=2--36 (35 IPs)} &&\\
\colrule
LSIC$^\text{b}$ & 0.206 & 0.170 &&\\
\end{tabular}
\end{ruledtabular}
\end{table}

\begin{table}[!htb]
\caption{\label{tabea}
The mean absolute error (in eV) of $\Delta$SCF electron affinities computed in various methods.}
\footnotetext{Reference [\onlinecite{doi:10.1063/1.2176608}].}
\begin{ruledtabular}
\begin{tabular}{lcccc}
&\multicolumn{2}{c}{SCAN}& PBE$^\text{a}$ & TPSS$^\text{a}$ \\ \cmidrule(lr){2-3} \cmidrule(lr){4-4} \cmidrule(lr){5-5}
\textrm{Method}&
\textrm{12 EAs}&
\textrm{20 EAs}&
\textrm{12 EAs}&
\textrm{12 EAs}
\\
\colrule
DFA\footnote{DFA results are based on $\Delta$SCF. The eigenvalue of an extra electron becomes positive (See text for details).}   & 0.115 & 0.148 & 0.13 & 0.05 \\
PZSIC        & 0.364 & 0.341 & 0.57 & 0.47 \\
OSIC ($k=1$) & 0.198 & 0.151 & 0.29 & 0.24 \\
OSIC ($k=2$) & 0.143 & 0.128 & 0.15 & 0.12 \\
OSIC ($k=3$) & 0.126 & 0.134 & 0.10 & 0.08 \\
OSIC ($k=4$) & 0.125 & 0.143 & & \\
\colrule
&\multicolumn{2}{c}{LSDA}&  &  \\ \cmidrule(lr){2-3} 
\textrm{Method}&
\textrm{12 EAs}&
\textrm{20 EAs}&&\\
\colrule
LSIC\footnote{Reference [\onlinecite{doi:10.1063/1.5129533}].} & 0.097 & 0.102 &&\\
\end{tabular}
\end{ruledtabular}
\end{table}

\begin{sidewaystable}[!htb]
\caption{\label{tabAE6}
The mean absolute and mean absolute percentage errors of AE6 set of molecules in various methods.} 
\footnotetext{Reference [\onlinecite{doi:10.1063/1.2176608}].}
\footnotetext{Reference [\onlinecite{doi:10.1063/1.5129533}].}
\begin{ruledtabular}
\begin{tabular}{lcccccccc}
        & \multicolumn{2}{c}{LSDA}& \multicolumn{2}{c}{SCAN}& \multicolumn{2}{c}{PBE$^\text{a}$}&\multicolumn{2}{c}{TPSS$^\text{a}$} \\ \cmidrule(lr){2-3} \cmidrule(lr){4-5} \cmidrule(lr){6-7} \cmidrule(lr){8-9}
       & MAE  & MAPE & MAE & MAPE & MAE & MAPE & MAE & MAPE \\
Method & (kcal/mol) & (\%) & (kcal/mol) & (\%) & (kcal/mol) & (\%) & (kcal/mol) & (\%) \\\hline
DFA	         & 74.26 (77.3)$^\text{a}$ & 15.93 (17.27)$^\text{a}$ &  2.85 & 1.15 & 15.5 & 4.43 & 5.9 & 2.43 \\
PZSIC	     & 57.97 (60.3)$^\text{a}$ & 9.37 (10.61)$^\text{a}$ & 26.52 & 7.35 & 17.0 & 5.54 & 34.7 & 9.29 \\
OSIC ($k=1$) & 18.93 (21.0)$^\text{a}$ & 3.96 (5.12)$^\text{a}$ & 8.67 & 2.78 & 12.6 & 4.21 & 9.9  & 4.01 \\
OSIC ($k=2$) & (8.6)$^\text{a}$ & (3.56)$^\text{a}$ & 4.86 & 2.12 & 16.0 & 4.76 & 11.3 & 4.09 \\
OSIC ($k=3$) & (7.2)$^\text{a}$ & (3.41)$^\text{a}$ & 4.18 & 1.93 & 17.2 & 5.20 & 12.4 & 4.10 \\
OSIC ($k=4$) & & & 4.10 & 1.85 & & & &\\
\colrule
OSIC ($z_\sigma =$ ELF, $k=1$)          & 23.21 & 4.85 & 9.04 & 2.68 & & & &\\
OSIC [$z_\sigma = 1-(2\beta)^2$, $k=1$] & 11.66 & 2.92 & 6.73 & 2.96 & & & &\\
\colrule
LSIC$^\text{b}$ & 9.95 & 3.20 & & &&&& \\
\end{tabular}
\end{ruledtabular}
\end{sidewaystable}

\begin{table}[!htb]
\caption{\label{tab:errBH6}
The mean and mean absolute errors (in kcal/mol) in barrier heights of BH6 set of molecules.}
\footnotetext{Reference [\onlinecite{doi:10.1063/1.2176608}].}
\footnotetext{Reference [\onlinecite{doi:10.1063/1.5129533}].}
\begin{ruledtabular}
\begin{tabular}{lcccccccc}
        & \multicolumn{2}{c}{LSDA}& \multicolumn{2}{c}{SCAN}& \multicolumn{2}{c}{PBE$^\text{a}$}&\multicolumn{2}{c}{TPSS$^\text{a}$} \\ \cmidrule(lr){2-3} \cmidrule(lr){4-5} \cmidrule(lr){6-7} \cmidrule(lr){8-9}
Method & ME & MAE & ME & MAE & ME & MAE & ME & MAE \\\hline
DFA	         & -17.62 (-17.9)$^\text{a}$ & 17.62 (17.9)$^\text{a}$ & -7.86 & 7.86 & -9.5 & 9.5 & -8.5 & 8.5 \\
PZSIC	     & -4.88 (-5.2)$^\text{a}$ & 4.88 (5.2)$^\text{a}$ & -0.81 &  2.96 & -0.1 & 4.2 & -0.2 & 5.7 \\
OSIC ($k=1$) & -2.95 (-3.2)$^\text{a}$ & 3.31 (3.5)$^\text{a}$ & -3.96 & 3.96 & -4.2 & 4.3 & -4.6 & 5.0 \\
OSIC ($k=2$) & (-2.8)$^\text{a}$ & (4.7)$^\text{a}$ & -5.45 & 5.45 & -6.5 & 6.5 & -6.8 & 6.8 \\
OSIC ($k=3$) & (-2.9)$^\text{a}$ & (5.7)$^\text{a}$ & -6.22 & 6.22 & -7.7 & 7.7 & -7.9 & 7.9 \\
OSIC ($k=4$) &&& -6.66 & 6.66 &&&&\\
\colrule
SOSIC ($k=4$) \footnote{Full PZSIC was applied on H-transfer FLOs (See text for details.)}  &&& -0.75 & 1.92 \\ 
\colrule
OSIC ($z_\sigma =$ELF, $k=1$)          & -3.08 & 3.15 & -4.25 & 4.25    \\
OSIC [$z_\sigma =1-(2\beta)^2$, $k=1$] & -1.88 & 2.28 & -4.87 & 4.87    \\
\colrule
LSIC$^\text{b}$ & 0.7 & 1.3 & & &&&&\\
\end{tabular}
\end{ruledtabular}
\end{table}

\begin{table}[!htb]
\caption{\label{tab:dissociation}
The mean absolute error (in kcal/mol) of SIE4$\times$4 and SIE11 sets of molecules. 
}
\footnotetext{Reference [\onlinecite{doi:10.1063/1.5129533}].}
\begin{ruledtabular}
\begin{tabular}{lcccc}
\textrm{Method}&
\textrm{SIE4$\times$4}&
\textrm{SIE11}&
\textrm{SIE11, 5 cationic}&
\textrm{SIE11, 6 neutral}\\
\colrule
SCAN & 17.9 & 10.1 & 10.4 & 9.9\\
PZSIC-SCAN  & 2.2 & 5.7  & 5.1 & 6.2  \\
OSIC-SCAN ($k=1$) & 2.9 & 14.7 & 6.5  & 21.5\\
OSIC-SCAN ($k=2$) & 5.2 & 13.4 & 7.6  & 18.2\\
OSIC-SCAN ($k=3$) & 6.5 & 7.4 & 7.7  & 7.0 \\
OSIC-SCAN ($k=4$) & 7.4 & 8.5 & 7.7  & 9.2   \\
\colrule
OSIC-LSDA ($z_\sigma =\tau^W/\tau$, $k=1$)  & 4.7 & 15.4 & 9.4 & 20.4 \\
OSIC-LSDA ($z_\sigma =$ELF, $k=1$)            & 7.3 & 18.0 & 11.8 & 23.1 \\
OSIC-LSDA [$z_\sigma =1-(2\beta)^2$, $k=1$] & 4.6 & 5.9  & 5.0  & 6.7\\
\colrule
OSIC-SCAN ($z_\sigma =$ELF, $k=1$)            & 5.0 & 7.3  & 6.4 & 7.9  \\
OSIC-SCAN [$z_\sigma =1-(2\beta)^2$, $k=1$] & 4.5 & 15.3 & 9.6 & 20.1 \\
\colrule
LSIC-LSDA$^\text{a}$ & 2.6 & 4.5 & 2.3 & 6.3 \\
\end{tabular}
\end{ruledtabular}
\end{table}

\begin{table}[!htb]
\caption{\label{tab:ehomo}
The mean absolute errors (in eV) in the highest occupied eigenvalues ($-\epsilon_{HO}$) for atoms hydrogen through argon and hydrogen through krypton. 
}
\begin{ruledtabular}
\begin{tabular}{lcc}
\textrm{Method}&
\textrm{$Z=1-18$ MAE}&
\textrm{$Z=1-36$ MAE}\\
\colrule
SCAN & 4.549 & 3.880 \\
PZSIC-SCAN  & 0.763 & 0.606 \\
OSIC-SCAN ($k=1$) & 1.051 & 1.045 \\
OSIC-SCAN ($k=2$) & 1.750 & 1.644 \\
OSIC-SCAN ($k=3$) & 2.151 & 1.981 \\
OSIC-SCAN ($k=4$) & 2.414 & 2.205 \\\hline
SOSIC-SCAN ($k=4$) & 0.754 &      \\ 
\end{tabular}
\end{ruledtabular}
\end{table}

\clearpage 
\bibliography{main}

\end{document}